\documentclass{article}
\usepackage{spconf,amsmath,amssymb,graphicx}
\usepackage[nolist, nohyperlinks]{acronym}
\usepackage{microtype}

\usepackage[backend=biber, style=ieee,maxbibnames=10]{biblatex}
\AtBeginBibliography{\footnotesize}

\begin{acronym}
\acro{stft}[STFT]{short-time Fourier transform}
\acro{istft}[iSTFT]{inverse short-time Fourier transform}
\acro{dnn}[DNN]{deep neural network}
\acro{pesq}[PESQ]{Perceptual Evaluation of Speech Quality}
\acro{polqa}[POLQA]{perceptual objectve listening quality analysis}
\acro{wpe}[WPE]{weighted prediction error}
\acro{psd}[PSD]{power spectral density}
\acro{rir}[RIR]{room impulse response}
\acro{snr}[SNR]{signal-to-noise ratio}
\acro{lstm}[LSTM]{long short-term memory}
\acro{polqa}[POLQA]{Perceptual Objectve Listening Quality Analysis}
\acro{sdr}[SDR]{signal-to-distortion ratio}
\acro{estoi}[ESTOI]{extended short-term objective intelligibility}
\acro{elr}[ELR]{early-to-late reverberation ratio}
\acro{tcn}[TCN]{temporal convolutional network}
\acro{rls}[RLS]{recursive least squares}
\acro{asr}[ASR]{automatic speech recognition}
\acro{ha}[HA]{hearing aid}
\acro{ci}[CI]{cochlear implant}
\acro{mac}[MAC]{multiply-and-accumulate}
\acro{vae}[VAE]{variational auto-encoder}
\acro{gan}[GAN]{generative adversarial network}
\acro{tf}[T-F]{time-frequency}
\acro{sde}[SDE]{stochastic differential equation}
\acro{drr}[DRR]{direct to reverberant ratio}
\acro{lsd}[LSD]{log spectral distance}
\acro{sisdr}[SI-SDR]{scale-invariant signal to distortion ratio}
\acro{mos}[MOS]{mean opinion score}
\acro{agc}[AGC]{automatic gain control}
\acro{vad}[VAD]{voice activity detection}
\end{acronym}

\title{Speech Signal Improvement Using Causal Generative Diffusion Models}
\name{Julius Richter, Simon Welker, Jean-Marie Lemercier, Bunlong Lay, Tal Peer, Timo Gerkmann  \thanks{\scriptsize This work has been funded by the German Research Foundation (DFG) in the transregio project Crossmodal Learning (TRR 169), DASHH (Data Science in Hamburg - HELMHOLTZ Graduate School for the Structure of Matter) with the Grant-No. HIDSS-0002, and the Federal Ministry for Economic Affairs and Climate Action, project 01MK20012S, AP380.}}
\address{Signal Processing (SP), Universität Hamburg, Germany}

\newcommand{\x}{\mathbf{x}}
\newcommand{\y}{\mathbf{y}}

\begin{document}
\ninept
\maketitle
\begin{abstract}
In this paper, we present a causal speech signal improvement system that is designed to handle different types of distortions. The method is based on a generative diffusion model which has been shown to work well in scenarios with missing data and non-linear corruptions. To guarantee causal processing, we modify the network architecture of our previous work and replace global normalization with causal adaptive gain control. We generate diverse training data containing a broad range of distortions. This work was performed in the context of an ``ICASSP Signal Processing Grand Challenge'' and submitted to the non-real-time track of the ``Speech Signal Improvement Challenge 2023'', where it was ranked fifth. 

\end{abstract}

\section{Introduction}
\label{sec:intro}
\vspace{-4px}

High-quality voice communication requires clear audio and natural-sounding speech. However, there are numerous factors that can degrade speech signals including background noise, room acoustics, transmission errors, limited bandwidth, and codec artifacts. Prior works on improving speech signals have typically studied each type of distortion separately. However, there has been some recent interest in developing universal approaches that address a broader range of distortions. These approaches typically use generative modeling, which works particularly well in scenarios with missing data and non-linear corruptions \cite{lemercier2023analysing}.

In this paper, we present our diffusion-based speech enhancement method submitted to the non-real-time track of the “Speech Signal Improvement Challenge 2023“ \cite{sig_challenge2023} as part of the “ICASSP Signal Processing Grand Challenges“. The proposed model extends our previous work \cite{richter2022speech}, incorporating significant modifications in the network architecture to meet the causality requirement and to output super wideband speech. Furthermore, we devise a data corruption approach to generate diverse training data resembling distortions observed in the blind data. Strong variations in loudness are compensated by using causal adaptive gain control. 

In the challenge's subjective test, our proposed method yields a final score of 0.445 using ITU-T P.863.2 and P.804 with \acp{mos} of 2.570 (Overall), 2.998 (Signal), 3.765 (Noise), 3.330 (Coloration), 3.674 (Discontinuity), 3.435 (Loudness), and 4.241 (Reverberation). In the subjective test based on ITU-T P.835 our method yields a final score of 0.495 with \acp{mos} of 2.842 (Overall), 3.119 (Signal), and 3.682 (Background). It is interesting to note that, as our approach is generative in nature, it is capable of achieving excellent performance for moderately distorted inputs, while it may generate phonetic confusions and insertions if the input distortions are too strong. 

\section{Proposed System}
\vspace{-4px}

\begin{figure}[tb]
\begin{minipage}[b]{1.0\linewidth}
  \centering
  \centerline{\includegraphics[scale=1.2]{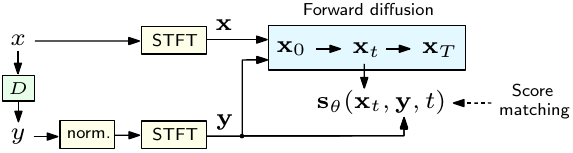}}
  \vspace{0.0cm}
  \centerline{(a) Training}\medskip
\end{minipage}

\begin{minipage}[b]{1.0\linewidth}
  \centering
  \centerline{\includegraphics[scale=1.2]{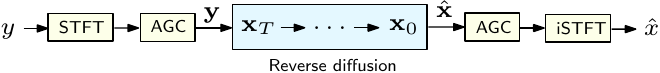}}
  \vspace{0.1cm}
  \centerline{(b) Inference}\medskip
\end{minipage}
\vspace{-2em}
\caption{Proposed system: (a) At training, a corruption model $D$ generates $y$ from clean speech $x$. The forward diffusion moves from the clean spectrogram $\x_0$ to the corrupted $\y$, while Gaussian noise is gradually added. The score model $\mathbf s_\theta$ is learned using score matching. (b) At inference, the data is normalized with causal adaptive gain control (AGC) and the reverse diffusion uses the trained score model.}
\label{fig:diagram}
\end{figure}

Fig. \ref{fig:diagram} shows an overview of our proposed system for speech signal improvement at training and inference time. The diffusion process (Sec. \ref{sec:diffusion_process}) is the core of the method and is accompanied by other processing blocks including the \ac{stft}, causal \ac{agc} (Sec. \ref{sec:agc}), and a data corruption model $D$ used to simulate various distortion types (Sec. \ref{sec:dataset}).

\subsection{Diffusion process}
\label{sec:diffusion_process}
\vspace{-2px}

The diffusion process $\{\x_t\}_{t=0}^T$ for speech enhancement is essentially the same as in our previous works \cite{richter2022speech, welker2022speech}, where $\x_t$ is the state of the process at time step $t$.
During training, the forward process moves from clean speech $\x_0:=\x$ to corrupted speech $\y$, while increasing amounts of Gaussian noise are gradually added.
At inference, the corresponding reverse process \cite{song2021sde} is used to progressively remove the corruption and therefore generate an estimate of $\x$ starting from $\x_T \sim \mathcal{N}_{\mathbb{C}}(\x_T; \y, \sigma_t^2)$. 
This reverse process involves the \emph{score function} of  $\x_t$, i.e., the gradient of its log-probability. Functioning as a prior for clean speech, it is unavailable at inference time and is thus approximated by a trained \ac{dnn} $\mathbf s_\theta$ called the \emph{score model}.
We make no further adaptations to this process as it is defined for every \ac{stft} bin independently, and is thus inherently causal as long as $\mathbf s_\theta$ is implemented with a causal network architecture.

\subsection{Network architecture}
\vspace{-2px}

We use a modified version of NCSN++ \cite{song2021sde} for score estimation. 
The network is an encoder-decoder architecture based on 2D convolutions, taking complex spectrograms $\mathbf x_t$ and $\mathbf y$ and the process time $t$ as input. Real and imaginary parts are considered as separate channels and the convolutions are performed over time and frequency.

We apply the following modifications to the architecture to meet the causality constraint:
\textbf{(1)} Padding in the 2D convolutions is modified so that the convolution along the time-dimension is causal;
\textbf{(2)} Batch normalization is replaced with cumulative group normalization, aggregating statistics recursively;
\textbf{(3)} Downsampling in the time dimension is performed with strided convolutions and corresponding upsampling with transposed strided convolutions. Up- and downsampling in the frequency dimension are realized with finite impulse response filters, as in \cite{richter2022speech}.   
\textbf{(4)} All attention layers as well as the progressive growing path are removed.

\subsection{Automatic gain control}
\label{sec:agc}
\vspace{-2px}

To match the unit-scale training condition of the score model, i.e., normalization of corrupted speech $y$ (see Fig.~\ref{fig:diagram}a), we use a causal \ac{agc} system before feeding the mixture to the diffusion process, and again after enhancement to maximize loudness, as part of the signal improvement task (see Fig. \ref{fig:diagram}b).
To this end, we recursively track the maximum value per magnitude frame averaged over the frequency bins to normalize the spectrogram in a causal manner. We start tracking when speech activity is first detected, using a voice activity detection method based on a causal speech presence probability estimator \cite{gerkmann2011_noise}. The speech probability is fed through an ideal low-pass filter to avoid having high-frequency noise bursts produce false positives. Voice activity is then assumed if the speech presence probability is higher than a threshold $\tau=0.8$, for a duration of $100$ms. When discovering a larger maximum than the previous one, we smooth the normalization with an exponential ramp going from the old value to the new one. 
Finally, to avoid clipping, we use a causal compressor from the \texttt{pedalboard} library\footnote{\scriptsize\texttt{https://github.com/spotify/pedalboard}}.

\section{Experimental Setup}
\vspace{-2px}

\subsection{Dataset}
\label{sec:dataset}
\vspace{-2px}

We use the VCTK corpus \cite{yamagishi2019cstr} as the clean speech dataset and resample all utterances from 48kHz to our processing sampling frequency 32kHz. 
Using the \texttt{audiomentations} library\footnote{\scriptsize\texttt{https://github.com/iver56/audiomentations}}, we simulate several corruptions observed in the blind data, namely stationary and non-stationary noise, reverberation, clipping, gain reduction, packet loss and lossy speech coding (the last two implemented by us). For each clean utterance, a random corruption chain is chosen among plausible candidates, e.g. $\{ \mathrm{Reverb} \rightarrow \mathrm{Noise} \rightarrow \mathrm{PacketLoss}  \}$, with the parameters of each corruption chosen randomly as well. We use the QUT corpus \cite{dean2010qut} as the environmental noise dataset and take room impulse responses from the DNS challenge \cite{dubey2022icassp} for reverberation.

\subsection{Hyperparameters and training configuration}
\vspace{-2px}

All processing is performed at $f_s=32$kHz and we upsample the processed files to the original $48$kHz frequency.
We use an \ac{stft} with a $638$-point Hann window and $160$-point hop, which guarantees the global latency to be below $20$ms, as we use purely causal processing. We set the lowest two frequency bins of the spectrogram to zero, to remove the DC offset and low frequency noise. 
The diffusion process hyperparameters are identical to those in \cite{richter2022speech}.
We train the score model $\mathbf{s}_\theta$ with the denoising score matching objective \cite{song2021sde} using the Adam optimizer with mini-batch size of 8 and exponential moving average over the parameters with a factor of $0.999$. Training takes around two days on two NVIDIA RTX A$6000$ GPUs.

\section{Results}
\vspace{-2px}

\begin{figure}[h]
\begin{minipage}[b]{0.31\linewidth}
  \centering
  \centerline{\includegraphics[width=\textwidth]{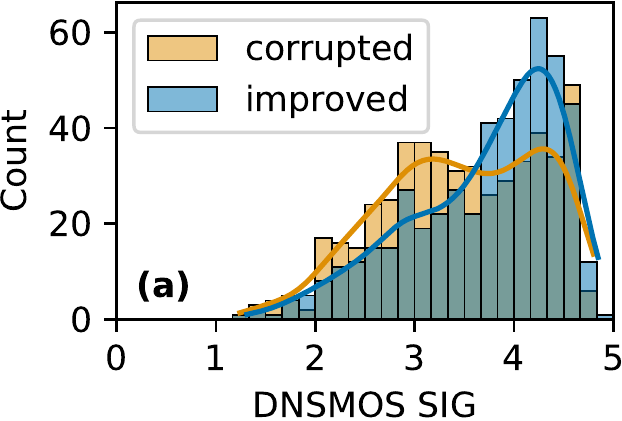}}
  \vspace{0.0cm}
\end{minipage}
\hfill
\begin{minipage}[b]{0.31\linewidth}
  \centering
  \centerline{\includegraphics[width=\textwidth]{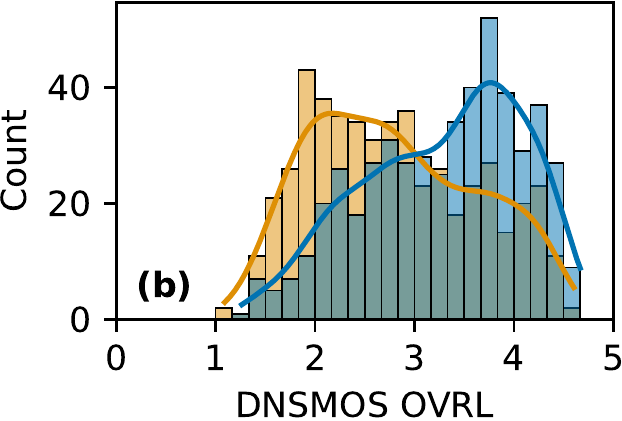}}
\end{minipage}
\hfill
\begin{minipage}[b]{0.31\linewidth}
  \centering
  \centerline{\includegraphics[width=\textwidth]{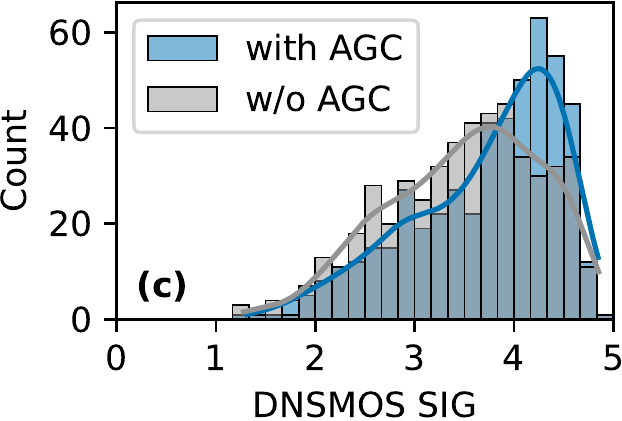}}
  \vspace{-0px}
\end{minipage}
\vspace{-2px}
\caption{Speech signal improvement results obtained with DNSMOS.}
\label{fig:results}
\end{figure}

In addition to the challenge's subjective evaluation \cite{sig_challenge2023}, we use DNSMOS P.835 \cite{reddy2022dnsmos} to evaluate our method on the 500 files in the blind set. 
Fig.~\ref{fig:results} depicts the histograms of DNSMOS scores: (a), (b) speech quality (SIG) and overall quality (OVRL) of the improved files compared to the corrupted files; (c) SIG with and without \ac{agc}.

Computational complexity: The network has about 55.7M parameters and the time to infer a frame on a CPU (Intel Core i7-7800X @ 3.50GHz) takes 0.89s. Please note that the model is designed to run on a GPU, on which the inference time is orders of magnitude faster (0.02 s$\,$/$\,$frame with an NVIDIA GeForce RTX 2080 Ti).

\section{Conclusion}
\vspace{-2px}

In this work, we have built upon our previous work on diffusion-based speech enhancement, with the novel contribution in making the system causal and training the model on different distortion types. The proposed method was ranked fifth in the non-real-time track of the ``Speech Signal Improvement Challenge 2023''.

\label{sec:refs}

\section{References}
\vspace{-2px}

\printbibliography[heading=none]

\end{document}